\documentclass[letter,scriptaddress,twocolumn, prl,showkeys]{revtex4}

	\usepackage{amsmath}
	\usepackage{makeidx}
	\usepackage{amsfonts}
	\usepackage[ansinew]{inputenc}
	\usepackage[usenames,dvipsnames]{pstricks}
	\usepackage{subfigure}
	\usepackage{epsfig}
	\usepackage{pst-grad} 
	\usepackage{pst-plot} 
	\usepackage[colorlinks,hyperindex]{hyperref}
	\makeatletter
\renewcommand{\@biblabel}[1]{#1.}
\makeatother

	\hypersetup
	{
		colorlinks,%
		citecolor=black,%
		linkcolor=black,%
		urlcolor=black,%
	}

\makeindex

\begin{document}
\title[Parasitic Signals in the Method of Recoil Nuclei]
{Analysis of Parasitic Signals in the Method of Recoil Nuclei Applied to Direct Observation of the $~^{229m}$Th Isomeric State}
\author{P.V.~Bilous}
\email{p.v.belous@gmail.com}
\author{L.P.~Yatsenko}
\affiliation{Institute of Physics, National Academy of Science of Ukraine}%
\address{Nauky Avenue 46, Kyiv 680028, Ukraine}

\setcounter{page}{1}

\begin{abstract}
We carry out necessary theoretical justifications for the method of recoil nuclei in application to direct observation of the $~^{229m}$Th isomeric state. We consider Cherenkov radiation, phosphorescence and fluorescence in the crystal plate which is used for collecting of thorium recoil nuclei and discuss the ways to avoid these parasitic signals revealing the $~^{229m}$Th decay photons.
\end{abstract}


\maketitle

\section{Introduction}

The isotope $~^{229}$Th is of significant interest due to predicted isomeric state with the energy astonishingly low for nuclear physics~\cite{FirstThExp}. It is considered to lie in a range of several~eV~\cite{35eV, 55eV, 76eV} typical for transitions in atomic electron shell, so this feature could allow to explore correlations between atomic and nuclear degrees of freedom. According to the latest indirect investigations this isomeric state has the energy $7.8 \pm 0.5$~eV and the half-life $\sim 5$~h~\cite{76eV}. By now, many fundamental theoretical investigations of processes in the atom $~^{229}$Th are made~\cite{Bridge_Karpeshin, Bridge_Porsev_3+, Bridge_Porsev_1+, Bridge_Porsev_Excit}. In addition such amazing applications as novel frequency standard with relative accuracy $10^{-19}$~\cite{Clock_10-19} and the first nuclear laser~\cite{NuclLaser} are suggested even in spite of the fact that there are no reliable proofs that the $~^{229m}$Th isomeric state really exists.

Actually, several attempts to find the state $~^{229m}$Th were made but with no success. Recently Zhao et al. reported the first observation of the $~^{229m}$Th decay photons~\cite{LosAlamos_Art}, but at once the result was called in question~\cite{LosAlamos_Comm}. By now the experiment has no unambiguous conclusions. In our opinion, the problem grows from the lack of theoretical justifications of the approach used therein. In this paper we fill this gap making necessary theoretical analysis.

The method used in Ref.~\cite{LosAlamos_Art} consists in implanting $~^{229}$Th nuclei recoiled from an $\alpha$-decaying $~^{233}$U ($T_{1/2} = 1.592 \cdot 10^5\text{ yr}$) into a plate prepared from material transparent in VUV ($\text{CaF}_2$, $\text{MgF}_2$, fused silica etc.). The isotope $~^{233}U$ decays $\sim 2\%$ branching to the state $~^{229m}$Th~\cite{2percIsom}. The UV photons which are emitted in result of the isomeric state decay are searched with photomultiplier (PMT). In the experiment Ref.~\cite{LosAlamos_Art} a uranium sample of the effective activity (that really takes part in implanting) at $\sim 170\text{ kBq}$ was used giving in authors' estimations $\sim 1000$~isomers/s in the crystal plate. For a few hours implantation period it lead to a signal on the order of 1 photon per second during measuring with the PMT.

The PMT fixes not only the $~^{229m}$Th radiation. First of all, phosphorescence in the crystal caused by $\alpha$-radiation in the implanting process may take place. In addition, as pointed out in Ref.~\cite{LosAlamos_Comm} if a uranium sample is not fresh enough there are accumulated daughter isotopes of $~^{233}$U implanting their recoil nuclei into the plate. These recoil nuclei in their turn decay giving $\alpha$- and $\beta$-particles leading to Cherenkov radiation and luminescence in the crystal. To make the method of recoil nuclei really consistent all these contributions must be analysed.

\section{Cherenkov radiation}

If the uranium sample is not fresh enough it has not only $~^{233}$U but its daughter nuclei, too, which give their recoil nuclei into the crystal plate. It may cause Cherenkov radiation. Actually, the threshold of this process in transparent medium with refractive index $n=1.5$ is equal to 170~keV for radiating electron's kinetic energy. There are fast-decaying nuclei of $~^{213}$Bi ($T_{1/2} = 45.59\text{ min}$) and $~^{209}$Pb ($T_{1/2} = 3.25\text{ h}$) in the decay chain of $~^{233}$U. They are $\beta$-radioactive with mean $\beta$-electron's energies 435 and 644~keV, respectively, so Cherenkov radiation may take place.

\begin{figure}
\includegraphics[width=8cm]{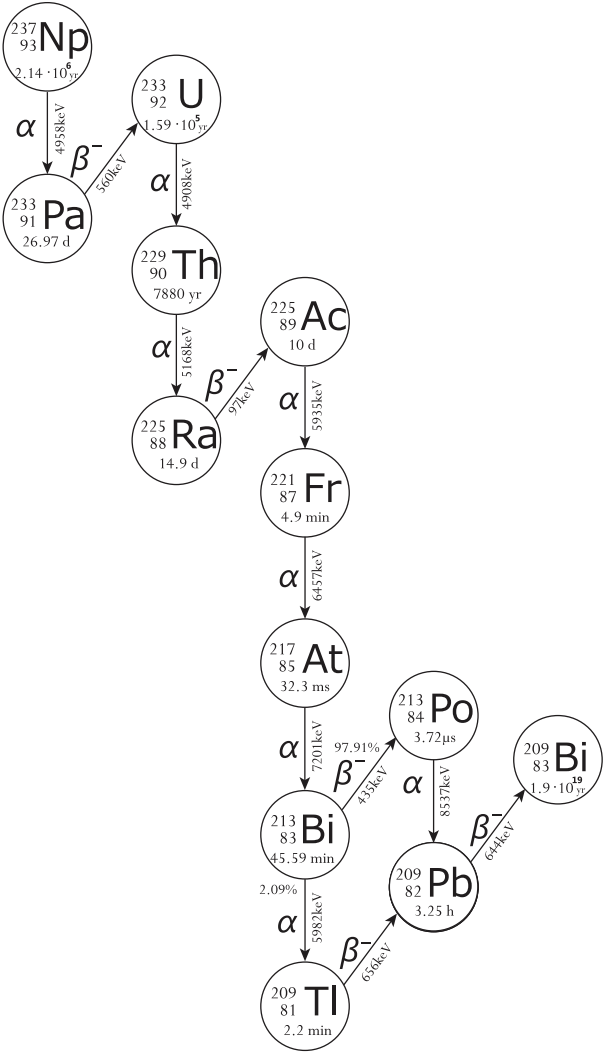}
\caption{\label{NeptSer}Neptunium series}
\end{figure}

The $~^{233}$U decay chain is the part of the neptunium series shown on Fig.~\ref{NeptSer}. There are the half-lives of the isotopes, the energies of their decays and the fractions of the decay branches. If $\beta$-decay takes place then the energy averaged over the spectrum of the $\beta$-electron is written. The $\beta$-radioactive nuclei with the energies exceeding the Cherenkov threshold are in the Table~\ref{tab:beta}.
The other $\beta$-emitters either have too small decay energy or don't belong to the main decay channel.

\begin{table}[b]
\noindent\caption{\label{tab:beta}$\beta$-active isotopes}\vskip3mm\tabcolsep4.2pt
\noindent{\footnotesize
\begin{tabular}{|c|c|c|c|}
\hline
Isotope & Half-life & $E_e$ (average) & The rate of the channel\\
\hline
$~^{213}$Bi & $45,59\text{ min}$ & $435 \text{ keV}$ & $97,8\%$\\
\hline
$~^{209}$Pb & $3,25\text{ h}$ & $644 \text{ keV}$ & $100\%$\\
\hline
\end{tabular}
}
\end{table}

We consider the age of the uranium sample to be much smaller than the half-life of $~^{229}$Th ($T_{Th} = 7400\text{ yr}$) and much greater than the half-lives of $~^{217}$At, $~^{213}$Po and the other nuclei after $~^{229}$Th. Also we note that $~^{229}$Th decays much faster than $~^{233}$U. Under these conditions the isotopic composition of the sample at the moment of the experiment may be considered as established and defined by the number of the $~^{229}$Th nuclei which estimated as $N_{Th}(t)=\lambda_U N_U t$, where $\lambda_U = \frac{\ln 2}{T_U}$ is the rate of the decay with the half-life $T_U$, $N_U$ denotes the number of the uranium nuclei and $t$ is the time of storing of the sample. The activity of the thorium is given by
\begin{equation}
A_{Th}(t) = \lambda_{Th} \cdot N_{Th}(t) = \frac{\ln 2}{T_{Th}}A_U t.
\end{equation}

The process of collecting of recoil nuclei in the crystal plate doesn't change total isotopic equilibrium (in the crystal plate and out of it). Further, let us assume that this process is carried out until the number of the isotopes in the plate becomes established. We consider that $~^{213}$Bi and $~^{209}$Pb in the plate constitute 1/2~part of their total number, because approximately the half has recoil impulses directed to the plate. Using it we deduce from the equilibrium condition (activities of each isotope are equal) the flux of $\beta$-electrons in the plate from each isotope causing Cherenkov radiation:
\begin{equation}\label{ABiAPb}
A_{Bi} = A_{Pb} = \frac{1}{2} A_{Th}(t) = \frac{1}{2} \cdot \frac{\ln 2}{T_{Th}}A_U t.
\end{equation}
In Ref.~\cite{LosAlamos_Art} the sample of the activity $A_U = 170\text{ kBq}$ was observed during $\sim 100$ days. From (\ref{ABiAPb}) we calculate that the nuclei which cause Cherenkov radiation has the activity $A_{Bi} = A_{Pb} \approx 2.2\text{ Bq}$.

Cherenkov loss of electron's energy per 1~cm is given by the formula~\cite{VavilCher}
\begin{equation}
\delta E = \frac{e^2}{c^2} \int_{\beta n(\omega)>1} \omega \Bigl[ 1- \frac{1}{\beta^2 n^2(\omega)}\Bigr] d\omega,
\end{equation}
where $e$ is the elementary charge, $c$ is the speed of light, $\beta$ is electron's speed in terms of $c$, integrating is carried out over frequency $\omega$ and $n(\omega)$ denotes spectral dependence of the refractive index. For $n(\omega) = n$ we immediately deduce number of irradiated photons per 1~cm in the range of photons' energies of $\Delta E$:
\begin{equation}\label{Ncheren}
\delta N = \frac{e^2}{\hbar^2 c^2} \Bigl[ 1- \frac{1}{\beta^2 n^2}\Bigr] \Delta E.
\end{equation}

\begin{figure}
\includegraphics[width=8cm]{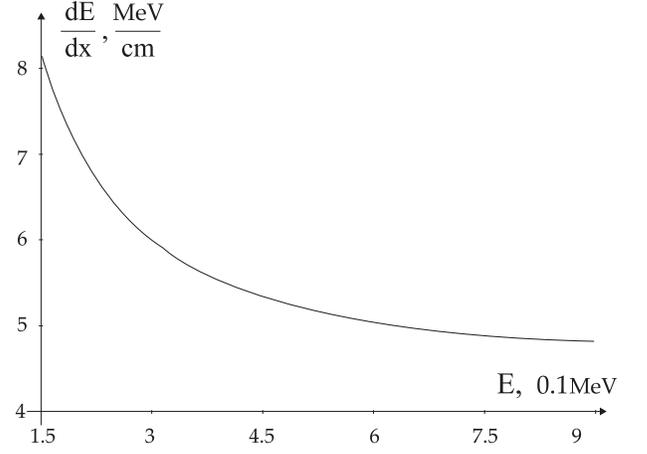}
\caption{\label{stop}Stopping force for electron in $\text{MgF}_2$}
\end{figure}

Using electronic database \cite{NIST} the loss of electron's energy (ionization + radiation) in the crystal $\text{MgF}_2$ in dependence of the energy magnitude was obtained (Fig. \ref{stop}). 
Total number of Cherenkov photons produced by an electron in the process of stopping from initial energy to the energy of the barrier $E_{min} = 170\text{ keV}$ equals
\begin{equation}
N = \int \frac{dN}{dE}dE = \int \frac{dN/dx}{dE/dx}dE,
\end{equation}
where integrating is carried out over the range of the energies from initial $E_e$ to the Cherenkov barrier $E_{min}$. Taking into account~(\ref{Ncheren}) we obtain
\begin{equation}
N = \frac{\alpha \Delta E}{\hbar c} I,
\end{equation}
\begin{equation}
I = \int \frac{1}{dE/dx}\Bigl[ 1 - \frac{1}{n^2(1 - \frac{1}{(1+E/m_ec^2)})} \Bigr]dE.
\end{equation}
Here $\alpha = \frac{e^2}{\hbar c} = 1/137$ is the fine-structure constant and as $\Delta E$ should be taken the pass band of the PMT which for Hamamatsu R8486 used in Ref. \citep{LosAlamos_Art} is equal to 6~eV. The results of numeric calculation are given in the Table~\ref{tab:cherenkres}.
\begin{table}[b]
\noindent\caption{\label{tab:cherenkres}Total number of photons per 1 electron}\vskip3mm\tabcolsep4.2pt
\noindent{\footnotesize
\begin{tabular}{|c|c|c|c|}
\hline
$\beta$-source & $E_e$ (average)& $I, \text{ }10^{-3}\text{ cm}$ & N\\
\hline
$~^{213}$Bi & $435 \text{ keV}$ & $10.9$ & $24$\\
\hline
$~^{209}$Pb & $644 \text{ keV}$ & $27.8$ & $62$\\
\hline
\end{tabular}
}
\end{table}

Total intensity of Cherenkov radiation is the sum over the $\beta$-electrons produced in the plate per second:
\begin{equation}\label{Atot}
A_{total} = \frac{1}{2}(A_{Bi} N_{Bi} + A_{Pb}N_{Pb}).
\end{equation}
The factor $1/2$ appears because the recoil nuclei stick near the surface of the crystal and a half of the $\beta$-electrons fly out from the crystal at once after producing. Using (\ref{Atot}) we obtain $A_{total} \approx 95 \text{ }\frac{\text{photons}}{\text{s}}$. In Ref.~\cite{LosAlamos_Art} $1\%$ optical coupling to the PMT sensor and $10\%$ PMT quantum efficiency were assumed. Under these assumptions the PMT must register $\sim 0.1$ Cherenkov photons per second. It is much smaller than expected useful signal in Ref. \citep{LosAlamos_Art}.

This rough estimation shows that one may neglect Cherenkov radiation caused by $\beta$-decaying daughter nuclei of $~^{233}$U if the sample was purified at the earliest 100~days ago. There is another way to understand whether Cherenkov radiation gives significant contribution or not. From~(\ref{Ncheren}) we find that Cherenkov photons are distributed uniformly by spectrum. This fact allows to verify whether this parasitic signal takes place or not. Actually, following rule holds: \textit{if investigated range of energies has a subrange with no PMT signal then there is no Cherenkov radiation at all}. It's especially useful if there are not only daughter nuclei of $~^{233}$U but some other sources of Cherenkov radiation.

The experiment Ref.~\cite{LosAlamos_Art} was carried out with $\text{MgF}_2$ plates with implanted $~^{229}$Th being explored with two PMTs: Hamamatsu R8487 (operating range 115 -- 195 nm) and Hamamatsu R8486 (operating range 115 -- 320 nm) analysing. The first PMT showed no signal, so in accordance with the rule we conclude that there was no Cherenkov radiation in contradiction with the comment Ref.~\cite{LosAlamos_Comm}.

\section{Phosphorescence}

Even if the uranium sample is quietly fresh and pure phosphorescence in the crystal may occur as the result of irradiation during implanting process. Authors of Ref.~\cite{LosAlamos_Art} used Mylar foil to catch the recoil nuclei and leave only phosphorescence. As the Mylar foil acts on the phosphorescence, too, its intensity was considered to decrease in scaling factor (1.15). But this influence in general case is not so trivial and, as correctly pointed in Ref.~\cite{LosAlamos_Comm}, an opportunity to exclude the phosphorescence with one scaling factor must be previously proved. We make this proof and in this way validate the method.

Electrons and holes created under the radiation thermalize and then are caught by the traps in the crystal. We know that filled traps associated with radiation defects light for micro- and nanoseconds~\cite{LuminNanosec}, giving fluorescence. Therefore the phosphorescence must be caused by traps of another origin. We consider them to be associated with contamination centres caused by prolonged storing of the crystal in the atmosphere. These traps stay filled for much longer times slowly recombining due to the following mechanism~\cite{LuminMechanism}. The electrons and the holes tunnel to each other under the crystal potential and then recombine giving in this way the phosphorescence. Its intensity may be found as
\begin{equation}
J = \int_0^{+\infty} f(r)w_T e^{-w_Tt} dr,
\end{equation}
where integrating is carried out over the distance between electrons and holes in the pairs (combinations of any electron and any hole), $t$ is time, $w_T = w_T(r)$ is tunneling probability, $f(r)$ denotes contribution of pairs with distance $r$ given by expression $f(r) \propto S(r) w (r,\nu)$ in which $S$ is the overlap integral of electron's and hole's $\psi$-functions in the pare and $w$ denotes distribution of the pares by distances. Note that only number $w$ depends on concentration of filled traps $\nu$, which in its turn depends on the dose of ionizing radiation. Under the assumption that the traps are scattered uniformly Poisson distribution gives
\begin{equation}
w(r, \nu) = 4\pi r^2 \nu \exp\Bigl(-\frac{4}{3}\pi r^3 \nu\Bigr).
\end{equation}
The number of filled traps $\nu$ is bounded by the number of contamination centres. We consider $\nu$ to be small enough for holding $\frac{4}{3}\pi r^3 \nu \ll 1$ for such $r$ that $S(r)$ significantly differs from zero, i.e. electron's and hole's $\psi$-functions overlap appreciably. Under this assumption the exponent may be replaced by 1 and we easily deduce $w(r, \nu) \propto \nu$ and $J = \nu F(t)$, where function $F$ doesn't depend on $\nu$. We see that variation of $\nu$ just acts on the magnitude but doesn't change qualitative temporal behaviour of the phosphorescence signal. The result is that \textit{if the crystal is contaminated not much, phosphorescence may be really excluded with one scaling factor}.

\section{Fluorescence}

There may be another appearance of $~^{233}$U daughter nuclei: their $\alpha$- and $\beta$-particles may cause excitation of the medium and further fluorescence. This parasitic signal is the most difficult for studying which must be carried out for each material of the crystal plate separately. In this paper we consider the effect for $\text{MgF}_2$ and extend then the conclusions over some other materials.

We start from fluorescence caused by $\beta$-particles. We know that it is selective by wavelength and in our case is significant at wavelength $\lambda > 225\text{ nm}$~\cite{BetaFluor}, consequently the excluding rule reliable for Cherenkov radiation give no information now. But the fluorescence at $\lambda > 225\text{ nm}$ turns out to be on the same order as Cherenkov radiation, so they appear only together. Analysing Ref.~\cite{BetaFluor} we conclude that $\beta$-fluorescence efficiency may be roughly estimated by the value
\begin{equation}
\varepsilon_f = 2 \text{ }\frac{\text{photons}}{10\text{ nm}\cdot \text{MeV} \cdot 4\pi}
\end{equation}
physical sense of which is the number of photons irradiated per $10\text{ nm}$ of the spectrum into a unit solid angle for the loss of electron's energy at $1\text{ MeV}$.

As for Cherenkov radiation, we have already deduced the number of the photons irradiated in the range 115 -- 320 nm per an electron decelerating from initial kinetic energy to the threshold of Cherenkov effect equal to 170 keV in $\text{MgF}_2$. The results are shown in Table~\ref{tab:cherenkres}. For subsequent estimations we use the value $n = 110 \text{ photons/MeV}$ as mean number of photons per lost energy 1 MeV. Now we easily evaluate an "efficiency" of Cherenkov radiation:
\begin{equation}
\varepsilon_c = 0.4 \text{ }\frac{\text{photons}}{10\text{ nm}\cdot \text{MeV} \cdot 4\pi}.
\end{equation}
We see that $\varepsilon_f$ and $\varepsilon_c$ differ just in 5 times, i.e. are comparable, so \textit{if there is no Cherenkov radiation at all then there is no $\beta$-fluorescence, too}. It works for almost all materials listed in Table II Ref.~\cite{BetaFluor}. It fails for $\text{CaF}_2$ and $\text{BaF}_2$ because their fluorescence efficiency is on the order greater and can not be compared with the efficiency of Cherenkov radiation. This fact complicates usage of these materials for nuclear recoil experiments.

If there is no Cherenkov radiation and $\beta$-fluorescence, and the phosphorescence is excluded, then only $\alpha$-fluorescence remains. To realize whether the resulting signal represents $\alpha$-fluorescence or something else one should consider its time dependence and compare it with time evolution of total $\alpha$-activity in the crystal plate looking for similar features. For example, as the authors of Ref.~\citep{LosAlamos_Comm} pointed $\alpha$-activity time dependence has a feature consisting in first decreasing because of the fast decay of $~^{221}$Fr (half-life 4.9~min), then building up over several days (half-lives of $~^{225}$Ra and $~^{225}$Ac).

Resuming obtained results we see that there was no Cherenkov radiation in the experiment Ref. \cite{LosAlamos_Art} and as $\text{MgF}_2$ was used there could be just little $\beta$-fluorescence. The phosphorescence was excluded with the scaling factor. The resulting signal has nothing like first decreasing and subsequent building during long time. Thus, we conclude that the authors of Ref. \cite{LosAlamos_Art} really might observe $~^{229m}$Th decay photons.

\section{Conclusion}

Applying the method of recoil nuclei one should use uranium sources fresh enough to exclude Cherenkov radiation and fluorescence in the crystal used for collecting of thorium recoil nuclei. We showed that one may neglect Cherenkov radiation caused by $\beta$-decaying daughter nuclei of $~^{233}$U if the sample was purified at the earliest 100~days ago. We deduced that if investigated range of energies has a subrange with no PMT signal then there is no Cherenkov radiation at all. Using these results we conclude that authors of Ref.~\cite{LosAlamos_Art} really did not deal with Cherenkov radiation. More than we have shown that the phosphorescence in this experiment was excluded correctly. Thus if there were no $~^{229m}$Th decay photons then resulting signal must represent only fluorescence. But under conditions in Ref.~\cite{LosAlamos_Art} time dependence of the signal isn't characteristic for fluorescence, so authors really might observe $~^{229m}$Th decay photons. We carried out necessary theoretical foundations for the method of recoil nuclei and we hope that other experiments with this method will be carried out and give necessary statistics for making final conclusion on the isomer state of $~^{229}$Th.

\end{document}